\documentclass[aps,prc,superscriptaddress,showpacs,amssymb,amsmath,amsfonts,floatfix]{revtex4}
\usepackage{graphicx}

\begin{document}

\title{Updated analysis of NN elastic scattering to 3~GeV}

\author{R.~A.~Arndt}
\author{W.~J.~Briscoe}
\author{I.~I.~Strakovsky}
\author{R.~L.~Workman}
\affiliation{Center for Nuclear Studies, Department of
        Physics, \\
        The George Washington University, Washington,
        D.C. 20052, U.S.A.}

\begin{abstract}
A partial-wave analysis of NN elastic scattering data
has been updated to include a number of recent measurements.
Experiments carried out at the Cooler Synchrotron (COSY)
by the EDDA Collaboration have had a significant impact 
above 1~GeV. Results are discussed in terms of the 
partial-wave and direct-reconstruction amplitudes. 
\end{abstract}

\maketitle

\section{Introduction}
\label{sec:intro}

In our previous analysis of nucleon-nucleon (NN) elastic 
scattering data~\cite{sp00}, the upper energy limit was 
extended from 2.5 to 3.0~GeV in the laboratory kinetic 
energy, T$_p$, in order to accomodate higher-energy 
proton-proton ($pp$) scattering measurements from SATURNE~II 
at Saclay and the EDDA Collaboration at COSY.  Comparisons 
were made with the direct amplitude reconstruction (DAR) 
and partial-wave analysis (PWA) study of Ref.~\cite{lehar1}, 
which gave amplitudes to 2.7~GeV. In some cases, where the 
DAR allowed 2 distinct solutions, our energy-dependent and 
single-energy results were seen to follow different sets 
of DAR amplitudes. With additional precise measurements, it 
was expected that these descrepancies would be removed. 
Such data have now appeared and our fits have changed 
substantially as a result. 

In the following, we will list those data recently added to 
our database. After a brief reminder of the differences 
between the PWA and DAR techniques, we will give our latest 
fit results. Some of the changes have been quite large, and 
these are discussed in light of the constraints imposed by 
the earlier Saclay DAR results. Finally, we summarize the 
status of the NN problem and consider what further work may 
be expected.

\section{Database}
\label{sec:expt}

The full database and a number of fits, from our group and 
others, are available through the on-line SAID facility~\cite{SAID}. 
Here we will concentrate only on those new measurements 
added since the SP00 (spring 2000) solution was published.
 
Table~\ref{tab:tbl1} lists recent (post SP00) contributions 
to the pp database. A major contribution has come from the 
EDDA collaboration. From this source we have added unpolarized, 
single-polarized, and double-polarized cross sections covering 
a wide energy range. Final A$_{yy}$ measurements from SATURNE~II 
have also been added. The PNPI group has provided P and D$_t$ 
at a single energy.

Table~\ref{tab:tbl2} lists post-SP00 contributions to the np 
database. Major contributions to the np (single and double) 
polarized data have come from PSI. The Uppsala facility has 
provided a detailed study of the angular shape of cross 
sections at 95~MeV. The IUCF group has done similar work at 
194~MeV. Low energy (below 20~MeV) cross sections have come
from Ohio University~\cite{bo02} and TUNL facility~\cite{wa01}, 
with an additional medium energy piece coming from 
JINR~\cite{sh04}.

\begin{table}[th]
\caption{Recent (since our previous 
         publication~\protect\cite{sp00}) pp elastic 
         scattering data up to 3~GeV. \label{tab:tbl1}}
\vspace{2mm}
\begin{tabular}{|c|c|c|c|c|c|}
\colrule
Observable          & Energy     & Angle    & Data & $\chi^2$ & Reference \\
                    &  (MeV)     & (deg)    &      &          &           \\
\colrule
d$\sigma$/d$\Omega$ & 240$-$2577 & 35$-$ 89 & 2888 & 3327     & \protect\cite{al04} \\
$P$                 & 437$-$2492 & 32$-$ 88 & 1131 & 1984     & \protect\cite{al05} \\
$A_{xx}$            & 481$-$2492 & 32$-$ 87 &  403 & 1197     & \protect\cite{ba05} \\
$A_{yy}$            & 481$-$2490 & 32$-$ 87 &  403 &  607     & \protect\cite{ba05} \\
$A_{zx}$            & 481$-$2490 & 32$-$ 87 &  403 & 1333     & \protect\cite{ba05} \\
$A_{yy}$            & 795$-$2795 & 47$-$105 &  477 &  671     & \protect\cite{al01} \\
$P$                 &1000        & 22$-$ 42 &    8 &   11     & \protect\cite{an04} \\
$D_t$               &1000        & 22$-$ 42 &    4 &   13     & \protect\cite{an04} \\
$A_{yy}$            &1795$-$2235 & 56$-$102 &  442 &  380     & \protect\cite{ag00} \\
\colrule
\end{tabular}
\end{table}
\begin{table}[th]
\caption{Recent (since our previous
         publication~\protect\cite{sp00}) np elastic
         scattering data up to 1.3~GeV. \label{tab:tbl2}}
\vspace{2mm}
\begin{tabular}{|c|c|c|c|c|c|}
\colrule
Observable          & Energy     & Angle    & Data & $\chi^2$ & Reference \\
                    &  (MeV)     & (deg)    &      &          &           \\
\colrule
$\Delta\sigma_L$    &   5$-$  20 &          &    6 &   11     & \protect\cite{wa01} \\
d$\sigma$/d$\Omega$ &  10        & 60$-$180 &    6 &    2     & \protect\cite{bo02} \\
$\Delta\sigma_T$    &  11$-$  17 &          &    3 &    7     & \protect\cite{wa01} \\
d$\sigma$/d$\Omega$ &  95        & 27$-$150 &   10 &    8     & \protect\cite{me04} \\
d$\sigma$/d$\Omega$ &  95        & 91$-$159 &    9 &   16     & \protect\cite{me05} \\
d$\sigma$/d$\Omega$ &  95        & 43$-$ 86 &    6 &   23     & \protect\cite{me06} \\
d$\sigma$/d$\Omega$ &  96        &152$-$175 &   11 &    8     & \protect\cite{kl02} \\
d$\sigma$/d$\Omega$ &  96        & 80$-$160 &    9 &   15     & \protect\cite{bl04} \\
d$\sigma$/d$\Omega$ &  96        & 20$-$ 76 &   12 &   25     & \protect\cite{jo05} \\
d$\sigma$/d$\Omega$ & 194        & 93$-$177 &   15 &   23     & \protect\cite{sa06} \\
$P$                 & 260$-$ 535 & 58$-$162 &  143 &  205     & \protect\cite{ar00} \\
$A_{yy}$            & 260$-$ 535 & 58$-$162 &  143 &  351     & \protect\cite{ar00} \\
$A_{zz}$            & 260$-$ 535 & 58$-$162 &  144 &  354     & \protect\cite{ar00} \\
$D$                 & 260$-$ 535 & 64$-$162 &   73 &  261     & \protect\cite{an00} \\
$D_t$               & 260$-$ 535 & 64$-$120 &   34 &   61     & \protect\cite{an00} \\
$A_t$               & 260$-$ 535 & 64$-$162 &   71 &   97     & \protect\cite{an00} \\
$R_t$               & 260$-$ 535 & 64$-$162 &   71 &   86     & \protect\cite{an00} \\
$N_{0s"sn}$         & 260$-$ 535 & 64$-$162 &   71 &   52     & \protect\cite{an00} \\
$N_{0s"kn}$         & 260$-$ 535 & 58$-$162 &   70 &   57     & \protect\cite{an00} \\
$N_{0nkk}$          & 260$-$ 535 &104$-$162 &   40 &   38     & \protect\cite{an00} \\
$D_{0s"0k}$         & 260$-$ 535 & 64$-$162 &   74 &  270     & \protect\cite{an00} \\
$P$                 & 284$-$ 550 &113$-$177 &  140 &  157     & \protect\cite{da02} \\
$\sigma^{tot}$      &1300        &          &    1 &    6     & \protect\cite{sh04} \\
\colrule
\end{tabular}
\end{table}

\section{Amplitude Analysis}
\label{sec:def}

Here, we are using the notation of Ref.~\cite{lehar1} and write
the scattering matrix, $M$, as
\begin{eqnarray}
M(\vec k_f , \vec k_i ) & = & {1\over 2} [ (a + b) +
       (a - b)~(\vec \sigma_1 \cdot \vec n)
              ~(\vec \sigma_2 \cdot \vec n)
   +   (c + d)~(\vec \sigma_1 \cdot \vec m )
              ~(\vec \sigma_2 \cdot \vec m ) \nonumber \\
 & + & (c - d)~(\vec \sigma_1 \cdot \vec l )
              ~(\vec \sigma_2 \cdot \vec l )
   +         e~(\vec \sigma_1 +\vec \sigma_2 )\cdot \vec n ] ,
\end{eqnarray}
where $\vec k_f$ and $ \vec k_i$ are the scattered
and incident momenta in the center-of-mass system,
and
\[ \vec n = { { \vec k_i \times \vec k_f }
        \over {|\vec k_i \times \vec k_f |} }
  \; , \;
   \vec l = { { \vec k_i + \vec k_f  }
        \over {|\vec k_i + \vec k_f | } }
  \; , \;
   \vec m = { { \vec k_f - \vec k_i  }
        \over {|\vec k_f - \vec k_i | } } .
 \]

Writing the scattering matrix in this form, any pp observable 
can be expressed in terms of the five complex amplitudes $a$ 
through $e$. If a sufficient number of independent observables 
are measured (precisely) at a given energy and angle, these 
amplitudes can be determined up to an overall undetermined 
phase. The advantage of this method is its model independence; 
nothing beyond the data is required to determine a solution. 
In addition, once the amplitudes are found, any further 
experimental quantity can be predicted at the energy-angle 
points of the DAR. There are, however, a number of 
disadvantages. The process gives amplitudes only at single 
energy-angle points, and no result is possible if an 
insufficient number of observables is available.

More standard is the PWA, which has observables constructed
from a series of amplitudes or phase shifts with allowed 
combinations of spin, angular momentum and isospin. This 
series must be cut off or augmented with a model for the 
high angular momentum states. It should be noted that, 
given a set of partial-wave amplitudes, the amplitudes $a$ 
through $e$ can be constructed, whereas the existence of 
amplitudes $a$ through $e$, at single energy-angle points, 
is insufficient to construct partial-wave amplitudes. 

At low energies, the PWA technique can generate a solution 
that is stable and requires less than a complete set of 
measurements. Inelasticity and a growing number of 
significant phase shifts make this method increasingly model 
dependent at higher energies.

\section{The Fit to 3~GeV}
\label{sec:fit}

Table~\ref{tab:tbl3} charts the evolution of our NN elastic 
scattering analyses. The present solution (SP07) and 
previously published analysis (SP00) are compared, in terms 
of fit $\chi^2$, in Table~\ref{tab:tbl4}. As in previous 
analyses, we have used the systematic uncertainty as an 
overall normalization factor for angular distributions.  
The description of this procedure is given in our recent 
$\pi$N PWA paper~\cite{piN}.  

\begin{figure*}[th]
\centerline{
\includegraphics[height=0.45\textwidth, angle=90]{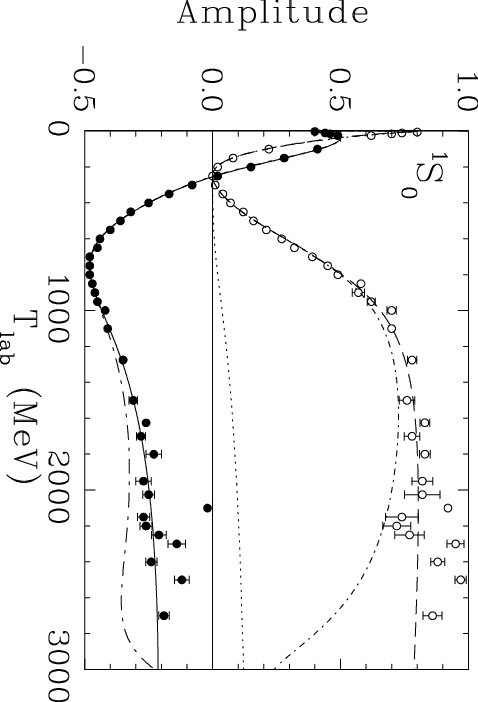}\hfill
\includegraphics[height=0.45\textwidth, angle=90]{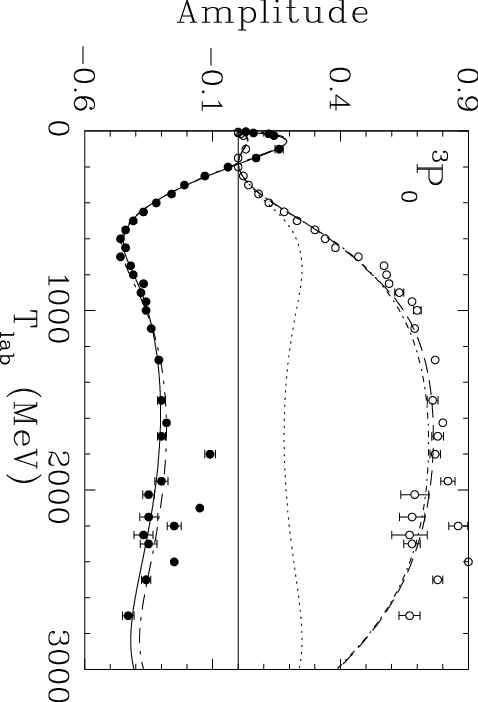}}
\centerline{
\includegraphics[height=0.45\textwidth, angle=90]{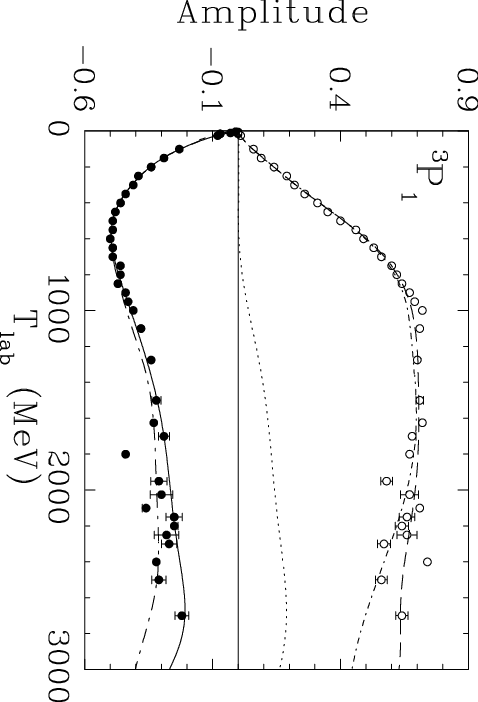}\hfill
\includegraphics[height=0.45\textwidth, angle=90]{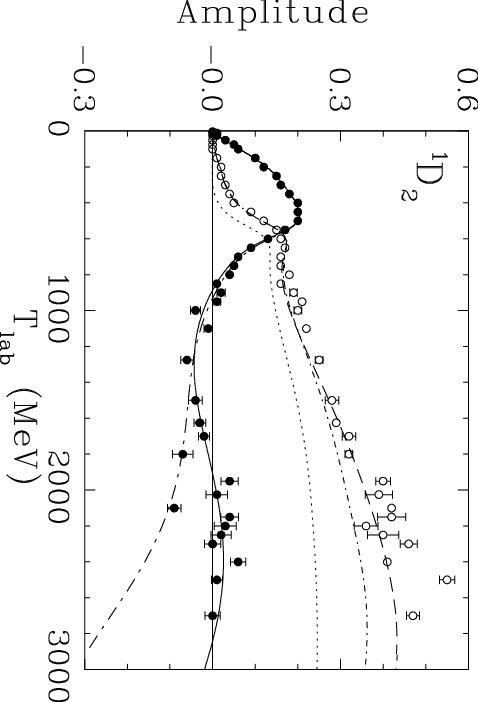}}
\centerline{
\includegraphics[height=0.45\textwidth, angle=90]{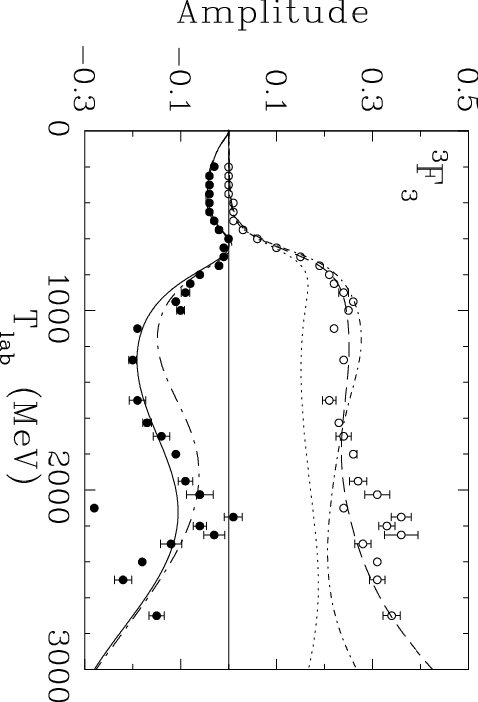}\hfill
\includegraphics[height=0.45\textwidth, angle=90]{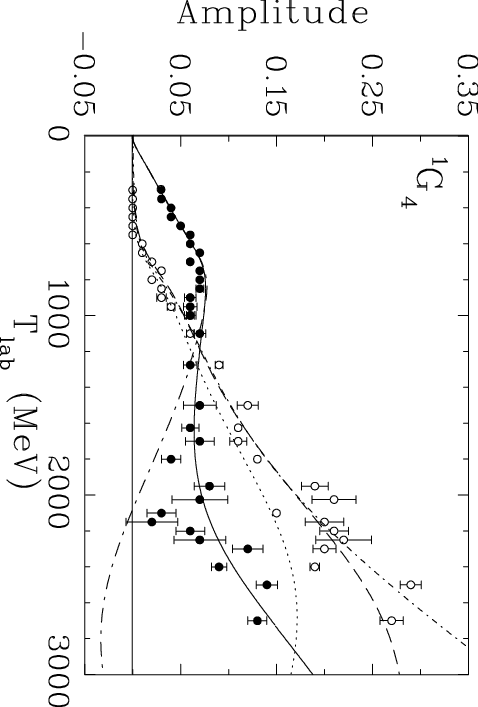}}
\centerline{
\includegraphics[height=0.45\textwidth, angle=90]{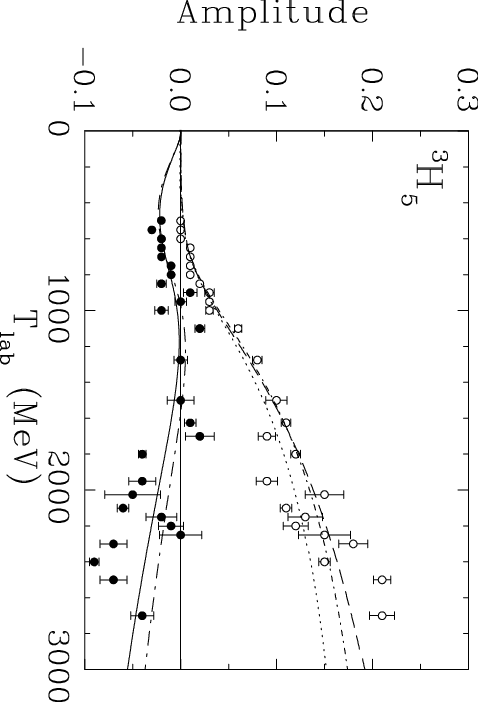}\hfill
}
\vspace{3mm}
\caption{Dominant isovector partial-wave amplitudes 
         from threshold to $T_p$ = 3~GeV.  Solid 
         (dashed) curves give the real (imaginary) 
         parts of amplitudes corresponding to the 
         recent SP07 solution.  The real (imaginary) 
         parts of single-energy solutions are plotted 
         as filled (open) circles.  The previous SP00 
         solution~\protect\cite{sp00} is plotted with
         dash-dotted (short dash-dotted) lines for the
         real (imaginary) parts.  The dotted curve 
         gives the unitarity limit $ImT - T^2 - 
         T^2_{sf}$ from SP07, where $T_{sf}$ is the
         spin-flip amplitude.  All amplitudes are 
         dimensionless. \label{fig:g1}}
\end{figure*}

\begin{figure*}[th]
\centerline{
\includegraphics[height=0.45\textwidth, angle=90]{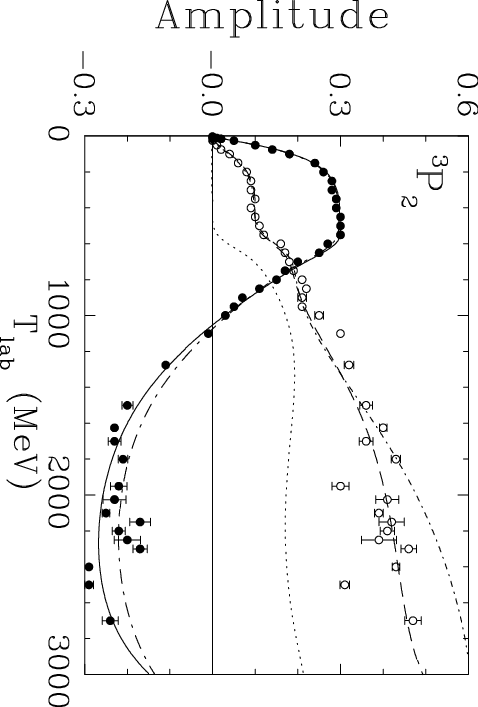}\hfill
\includegraphics[height=0.45\textwidth, angle=90]{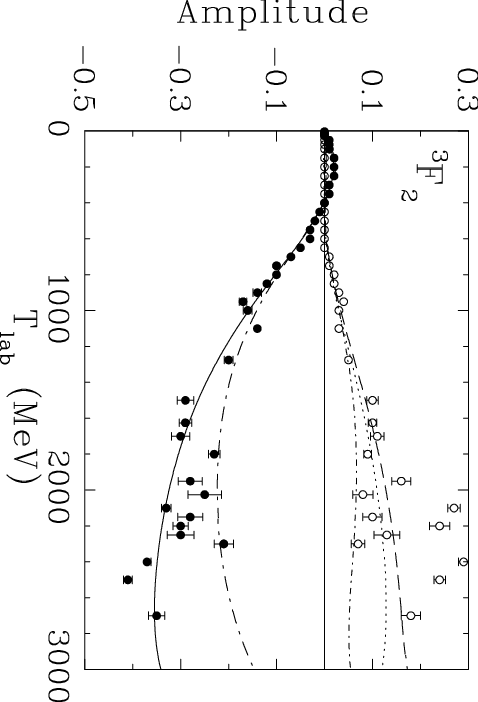}}
\centerline{
\includegraphics[height=0.45\textwidth, angle=90]{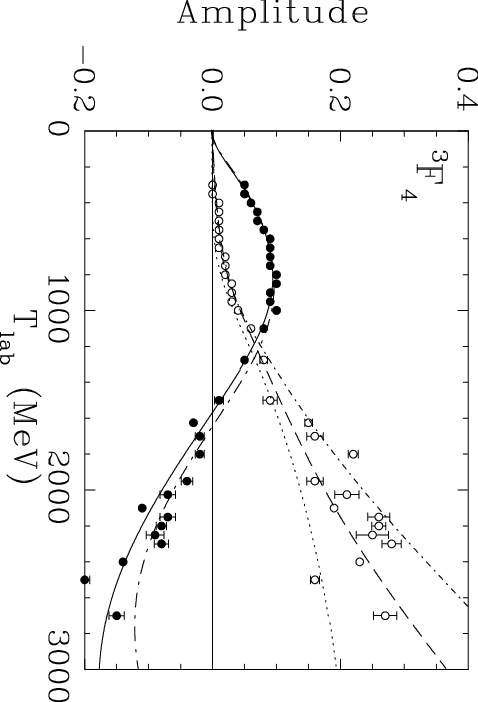}\hfill
\includegraphics[height=0.45\textwidth, angle=90]{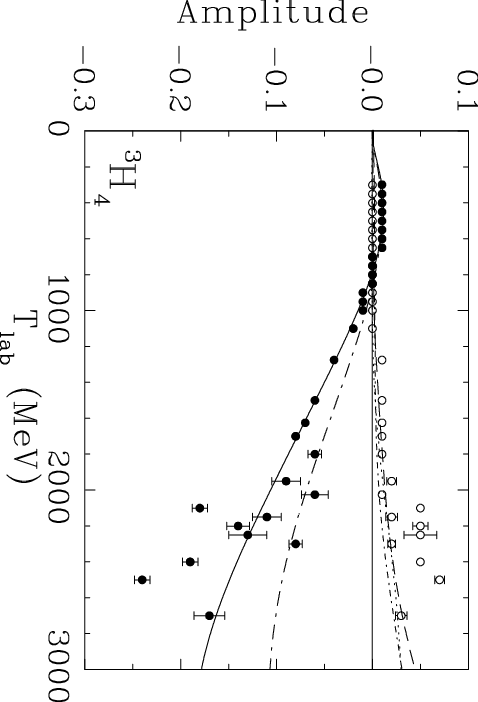}}
\centerline{
\includegraphics[height=0.45\textwidth, angle=90]{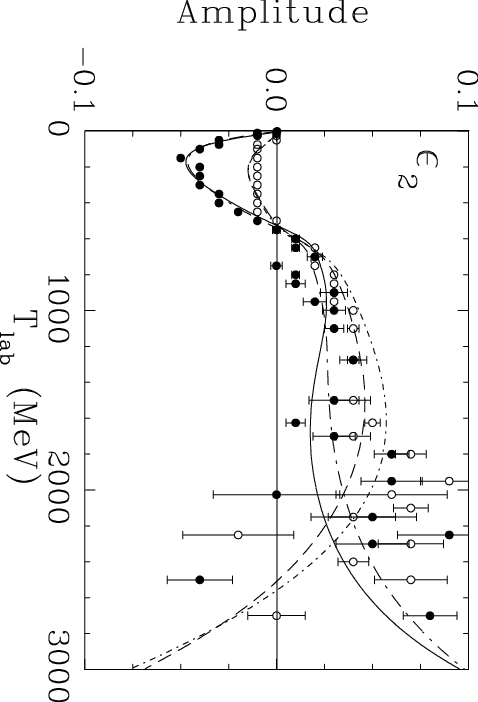}\hfill
\includegraphics[height=0.45\textwidth, angle=90]{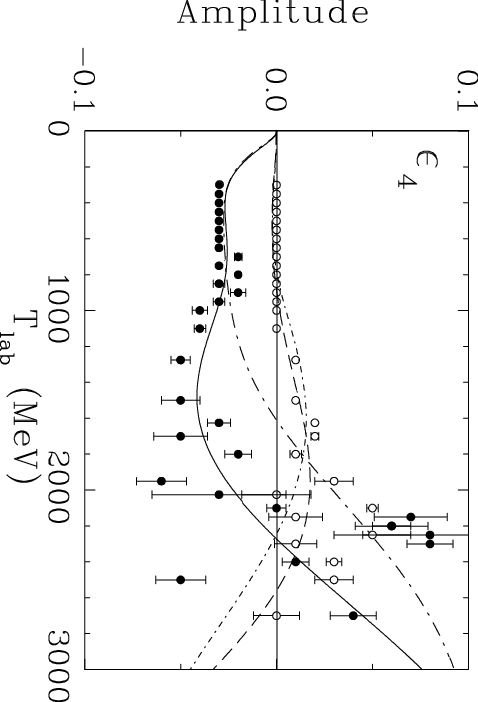}}
\vspace{3mm}
\caption{Notation as in Fig.~\protect\ref{fig:g1}. 
         \label{fig:g2}}
\end{figure*}

\begin{figure*}[th]
\centerline{
\includegraphics[height=0.45\textwidth, angle=90]{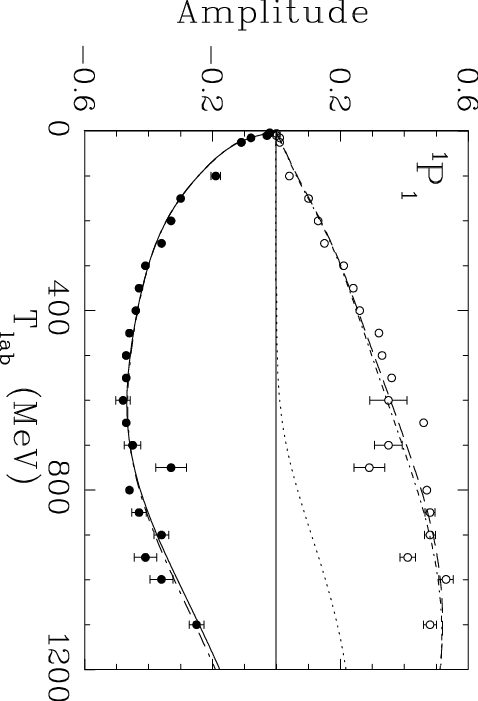}\hfill
\includegraphics[height=0.45\textwidth, angle=90]{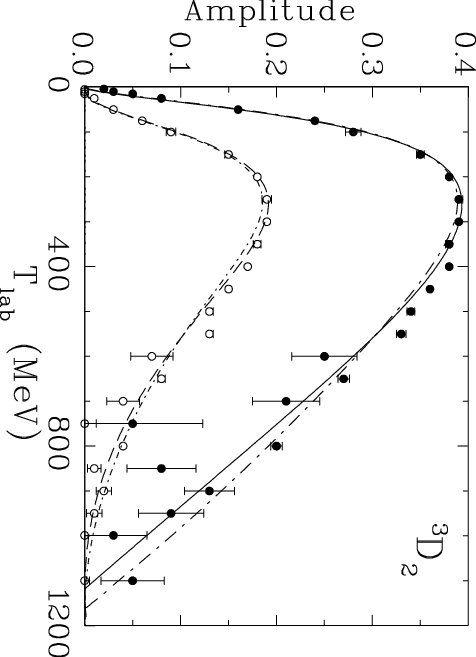}}
\centerline{
\includegraphics[height=0.45\textwidth, angle=90]{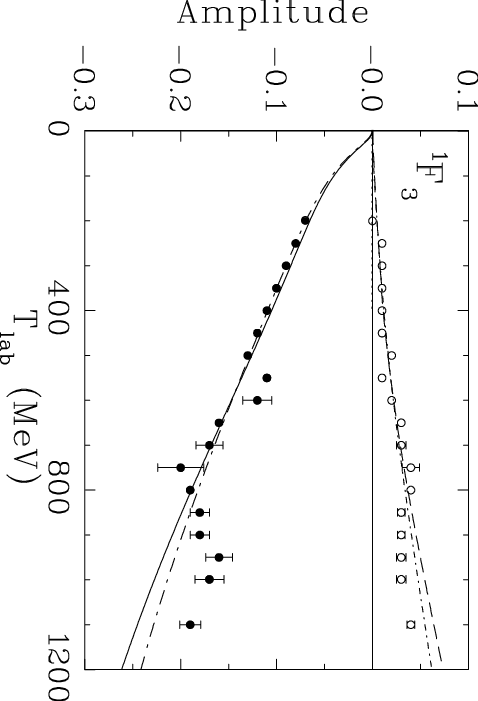}\hfill
\includegraphics[height=0.45\textwidth, angle=90]{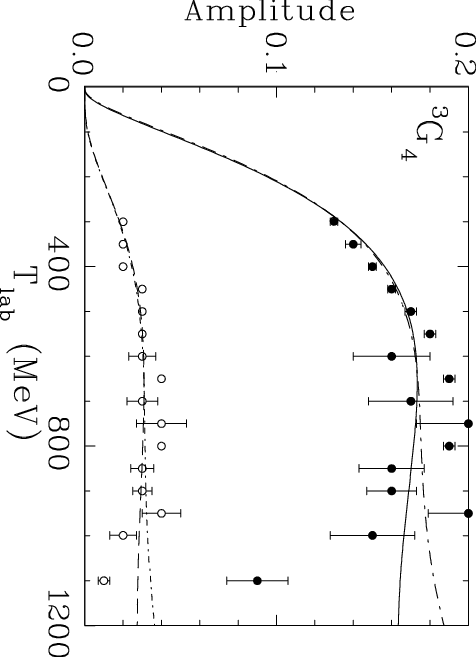}}
\vspace{3mm}
\caption{Dominant isoscalar partial-wave amplitudes
         from threshold to $T_p$ = 1.2~GeV.  
         Notation as in Fig.~\protect\ref{fig:g1}.
         \label{fig:g3}}
\end{figure*}

\begin{figure*}[th]
\centerline{
\includegraphics[height=0.45\textwidth, angle=90]{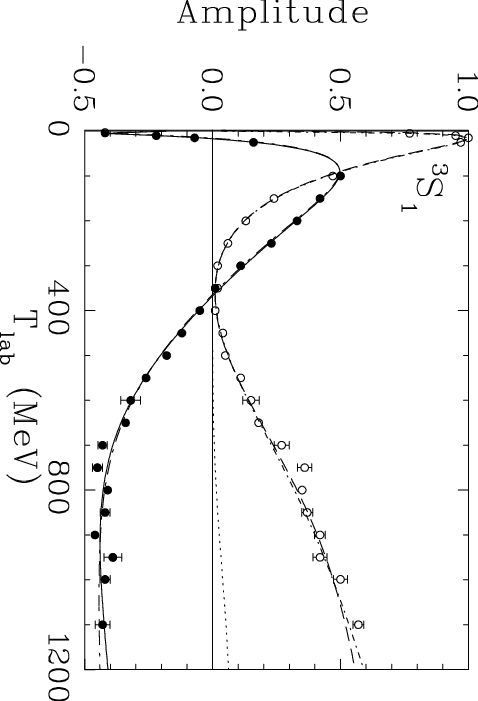}\hfill
\includegraphics[height=0.45\textwidth, angle=90]{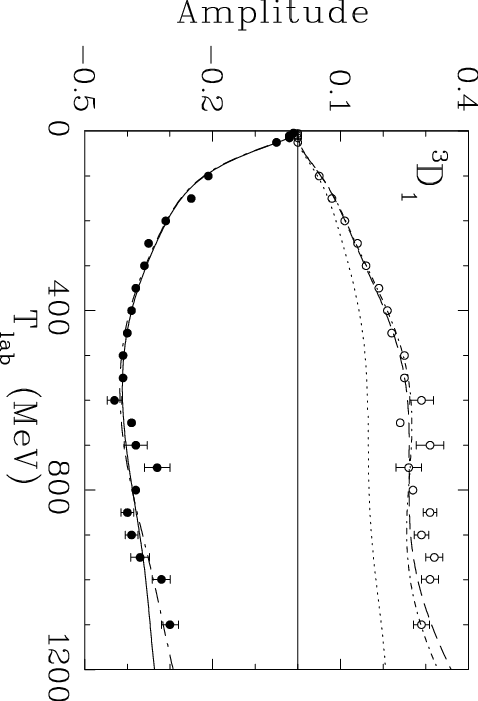}}
\centerline{
\includegraphics[height=0.45\textwidth, angle=90]{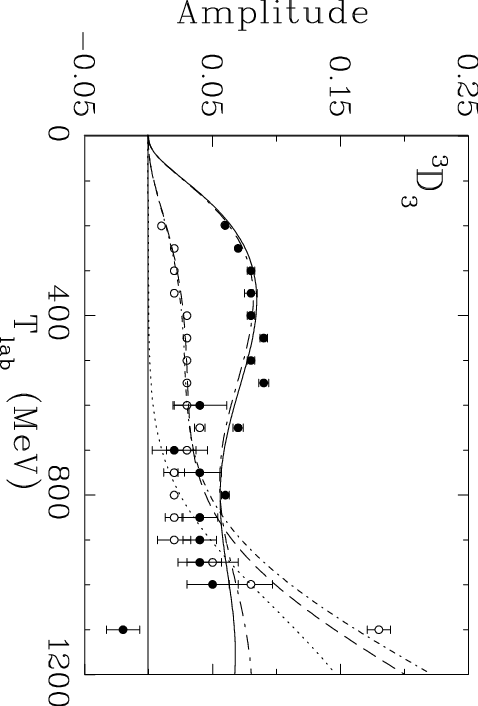}\hfill
\includegraphics[height=0.45\textwidth, angle=90]{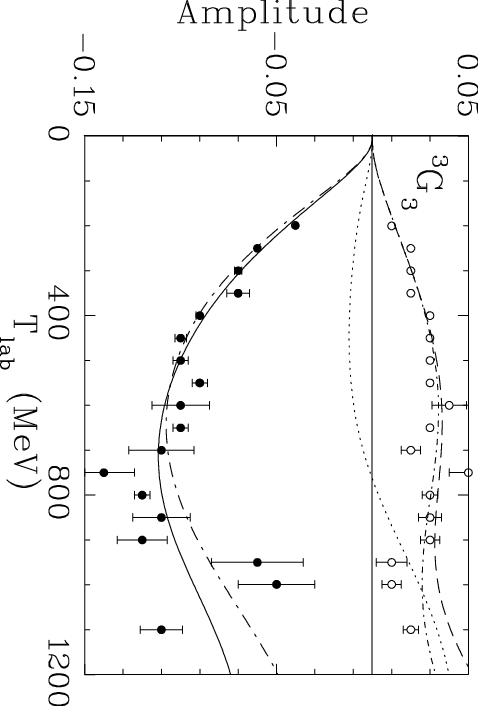}}
\centerline{
\includegraphics[height=0.45\textwidth, angle=90]{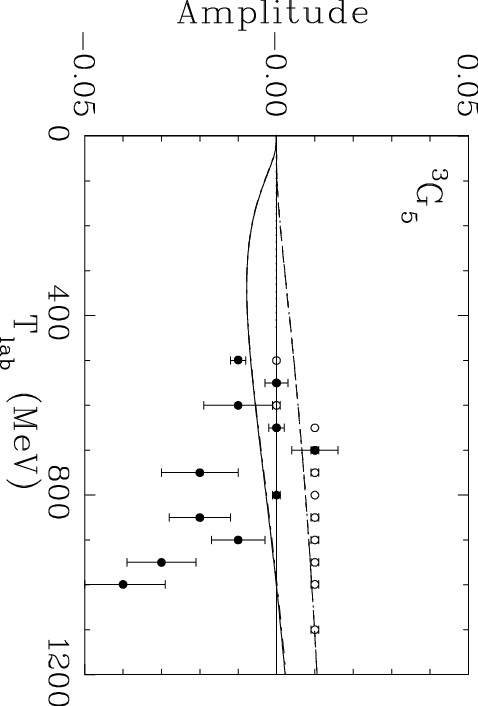}\hfill
\includegraphics[height=0.45\textwidth, angle=90]{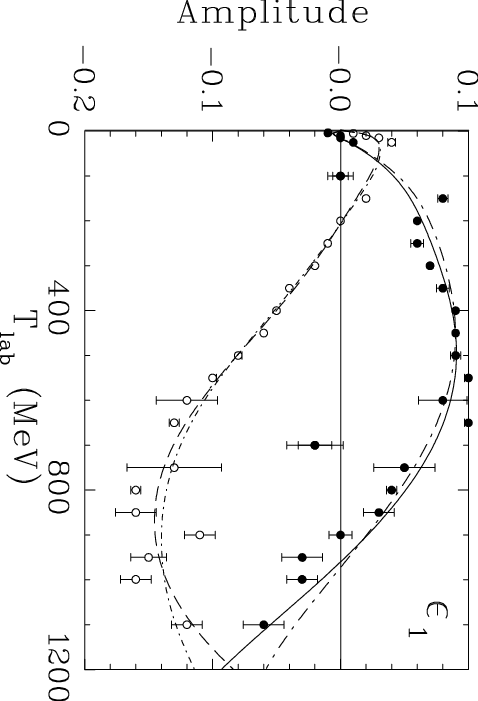}}
\centerline{
\includegraphics[height=0.45\textwidth, angle=90]{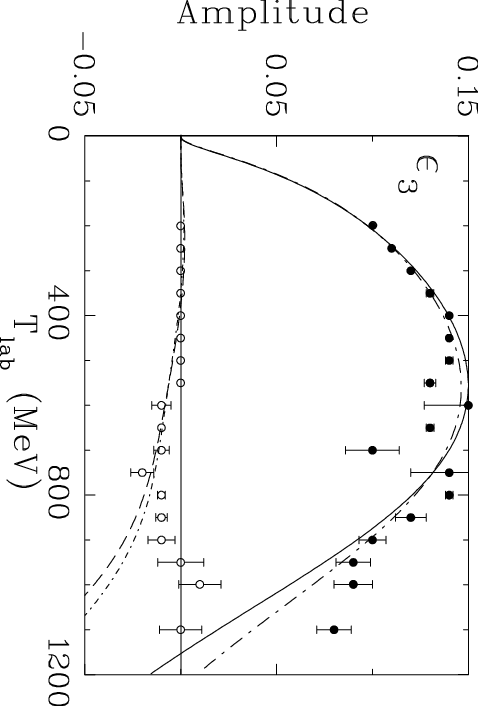}\hfill
\includegraphics[height=0.45\textwidth, angle=90]{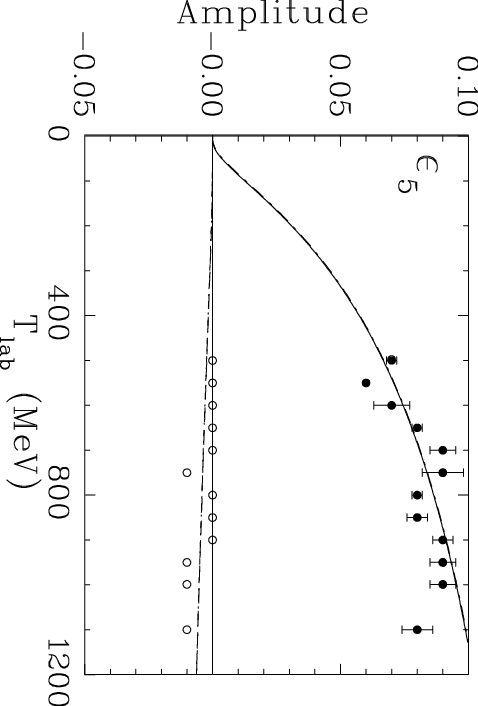}}
\vspace{3mm}
\caption{Notation as in Fig.~\protect\ref{fig:g3}. 
         \label{fig:g4}}
\end{figure*}

Below 1~GeV, where the $\chi^2$/data is near unity, the PWA 
solution has changed little. However, above this energy, 
qualitative changes can be seen in some amplitudes - in 
particular, the $^1D_2$ and $^1G_4$.  The dominant isovector 
partial-wave amplitudes are compared in Figs.~\ref{fig:g1} 
and \ref{fig:g2}.  No similarly large changes are evident 
in the isoscalar waves, which extend to only 1.3~GeV. These 
amplitudes are displayed in Figs.~\ref{fig:g3} and 
\ref{fig:g4}.

\begin{figure*}[th]
\centerline{
\includegraphics[height=0.45\textwidth, angle=90]{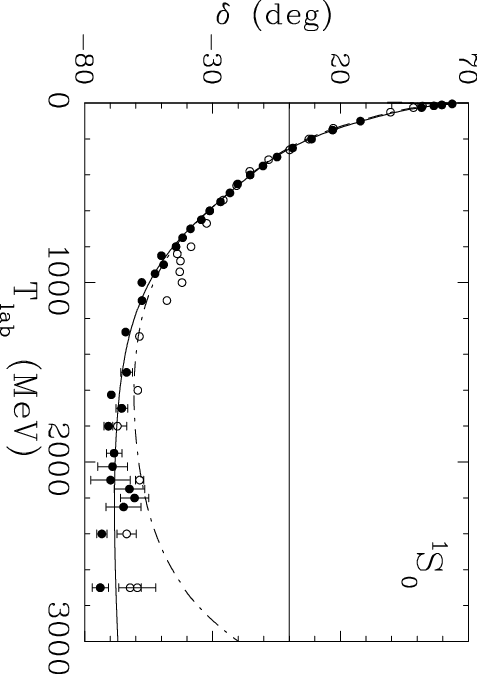}\hfill
\includegraphics[height=0.45\textwidth, angle=90]{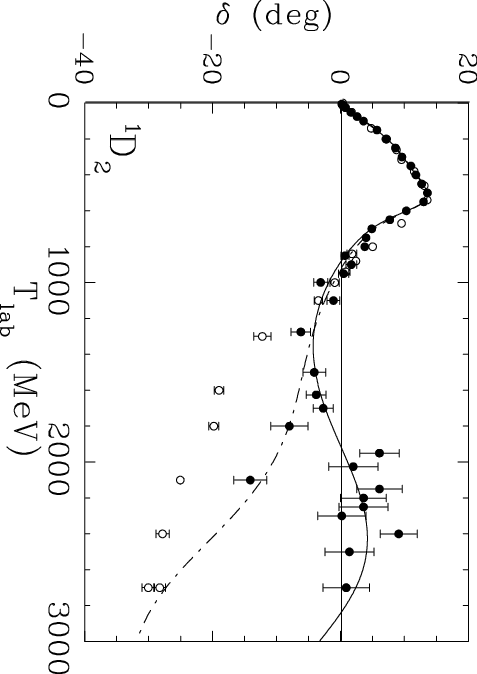}}
\centerline{
\includegraphics[height=0.45\textwidth, angle=90]{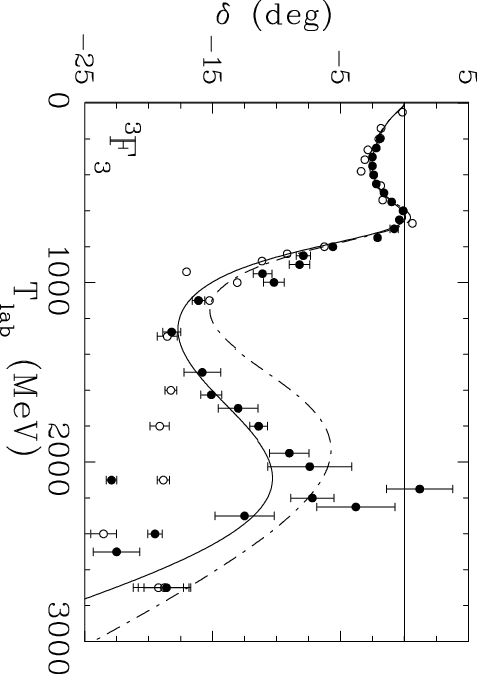}\hfill
\includegraphics[height=0.45\textwidth, angle=90]{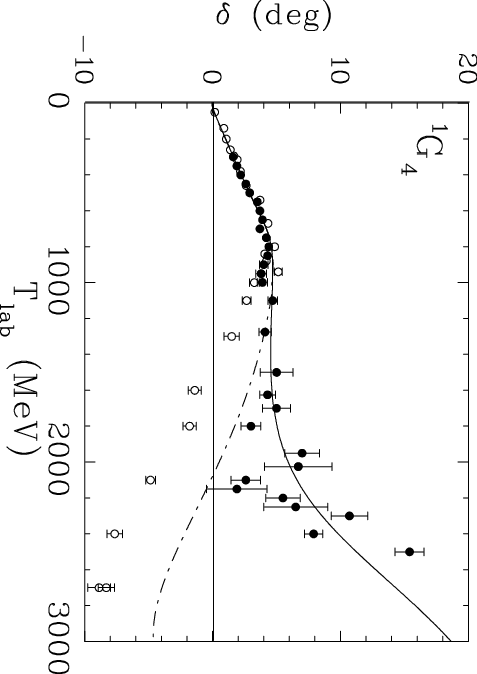}}
\vspace{3mm}
\caption{Phase-shift parameters for dominant isovector 
         partial-wave amplitudes from threshold to 
         $T_p$ = 3~GeV.  The SP07 and 
         SP00~\protect\cite{sp00} solutions are
         plotted as solid and dash-dotted curves,
         respectively.  The GW single-energy 
         solutions and those from 
         Saclay~\protect\cite{lehar1,lehar2} are 
         given by filled and open circles, 
         respectively. \label{fig:g5}}
\end{figure*}

Isovector phase shift parameters are given for the present 
energy-dependent and single-energy solutions, the 
energy-dependent SP00 solution, and single-energy Saclay 
analysis~\cite{lehar1,lehar2} in Fig.~\ref{fig:g5}.  Note 
that a qualitative agreement between the Saclay and SP00 
results for $^1D_2$ and $^1G_4$ is absent in SP07.

\begin{figure*}[th]
\centerline{
\includegraphics[height=0.75\textwidth, angle=90]{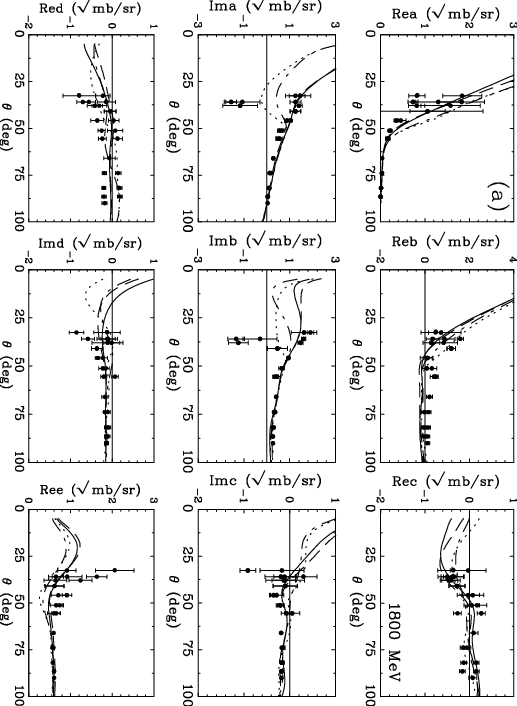}}
\centerline{
\includegraphics[height=0.75\textwidth, angle=90]{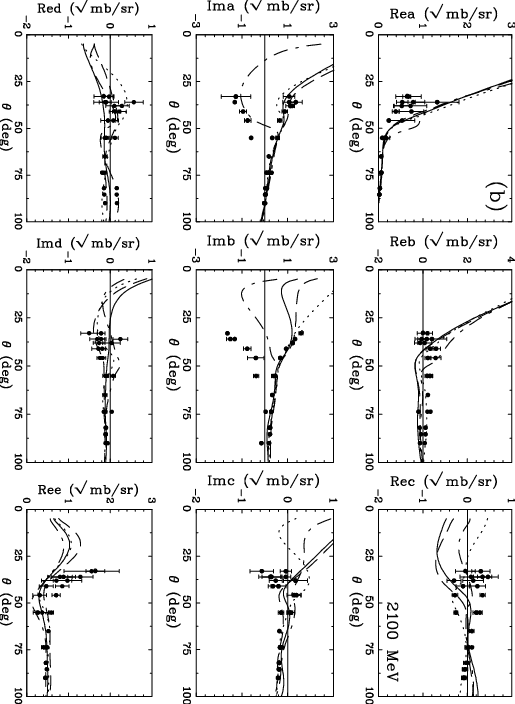}}
\vspace{10mm}
\caption{Direct-reconstruction amplitudes $a$ to $e$ 
         for $pp$ elastic scattering at (a) T$_p$ = 
         1.8~GeV and (b) 2.1~GeV as a function of 
         c.m. scattering angle.  The real and 
         imaginary parts of 
         amplitudes~\protect\cite{lehar1} are shown.
         Our SP07 (single-energy) solution is
         plotted with solid (dashed) lines.  Our 
         previous SP00 
         (single-energy)~\protect\cite{sp00}
         results are shown with dash-dotted (dotted)
         lines. \label{fig:g6}}
\end{figure*}

\begin{figure*}[th]
\centerline{
\includegraphics[height=0.75\textwidth, angle=90]{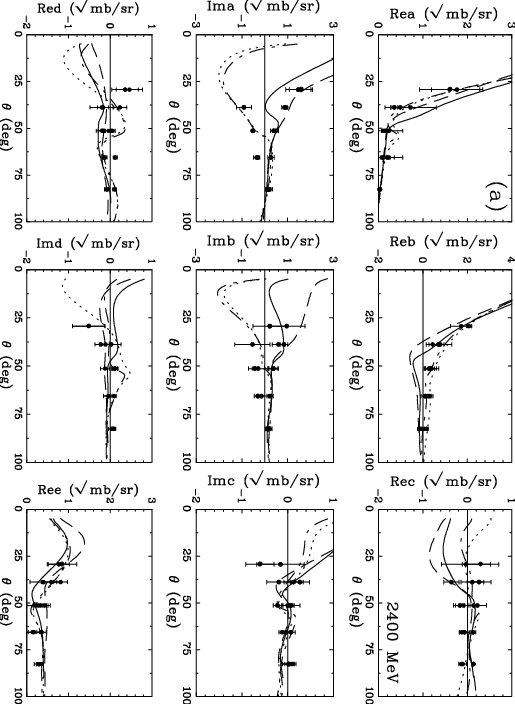}}
\centerline{
\includegraphics[height=0.75\textwidth, angle=90]{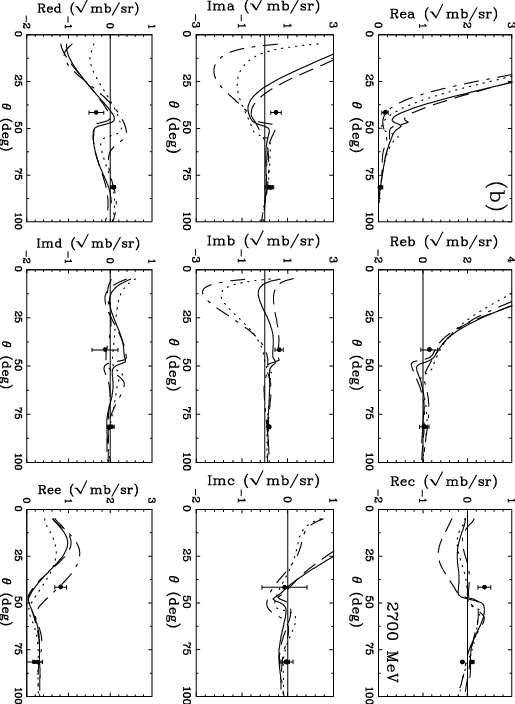}}
\vspace{10mm}
\caption{Direct-reconstruction amplitudes $a$ to $e$
         amplitudes for $pp$ elastic scattering at 
         (a) T$_p$ = 2.4~GeV and (b) 2.7~GeV as a 
         function of c.m. scattering angle.
         Notation as in Fig.~\protect\ref{fig:g6}.
         \label{fig:g7}}
\end{figure*}

Significant changes are also evident in comparisons with 
the Saclay DAR amplitudes. These are plotted in 
Figs.~\ref{fig:g6} and \ref{fig:g7}.  Here we have compared 
both the energy-dependent and single-energy results of SP00 
and SP07 with the Saclay values. The Saclay results, for 
some amplitudes, show two branches for the DAR. In the SP00 
publication, we noted that our single-energy and 
energy-dependent solutions were choosing different branches, 
particularly for the imaginary parts of $a$ and $b$. This 
discrepancy has largely disappeared in SP07. Both the 
energy-dependent and single-energy curves now follow a 
single branch of DAR results. 

\begin{figure*}[th]
\centerline{
\includegraphics[height=0.3\textwidth, angle=90]{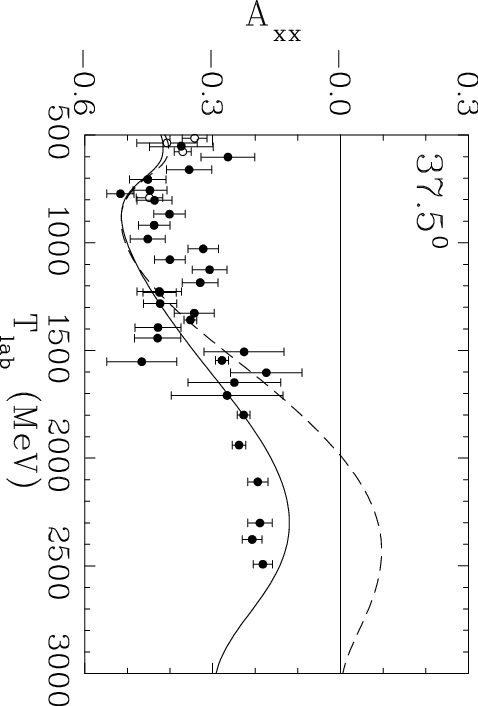}\hfill
\includegraphics[height=0.3\textwidth, angle=90]{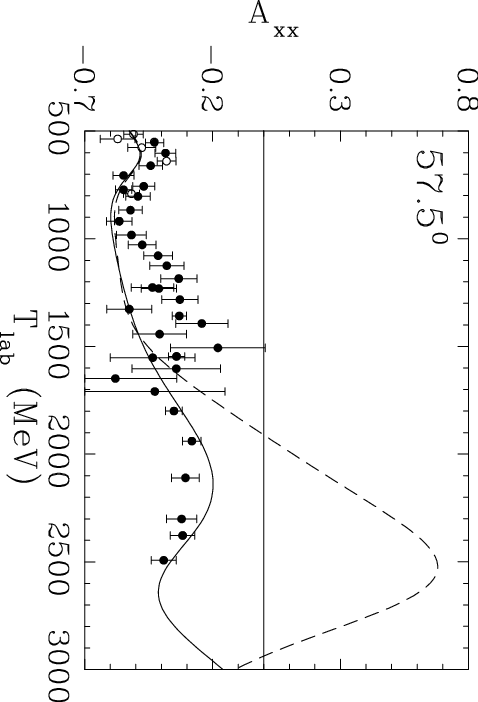}\hfill
\includegraphics[height=0.3\textwidth, angle=90]{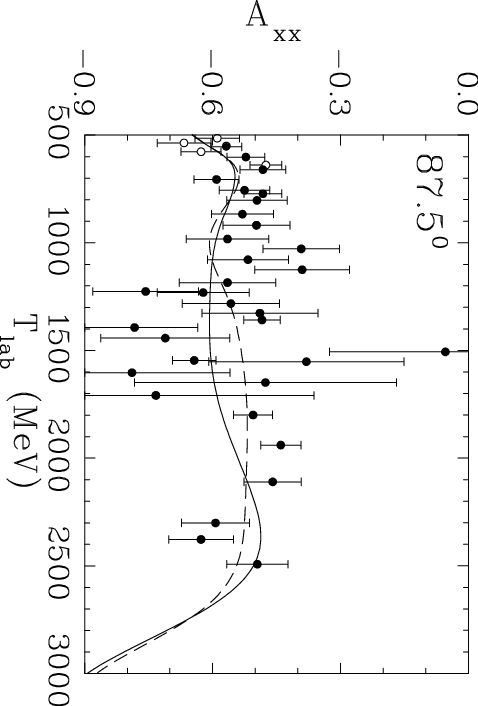}}
\vspace{10mm}
\caption{Excitation function A$_{xx}$ for $pp$ 
         elastic scattering at three c.m. 
         scattering angles.  The EDDA 
         Collaboration data (filled circles) are 
         from~\protect\cite{ba05}.  Other previous 
         measurements (for references see SAID 
         database~\protect\cite{SAID}) within a 5 
         degree bin are shown as open circles.
         The SP07 (SP00~\protect\cite{sp00}) 
         solution is plotted as a solid (dashed) 
         curve. \label{fig:g8}}
\end{figure*}

\begin{figure*}[th]
\centerline{
\includegraphics[height=0.3\textwidth, angle=90]{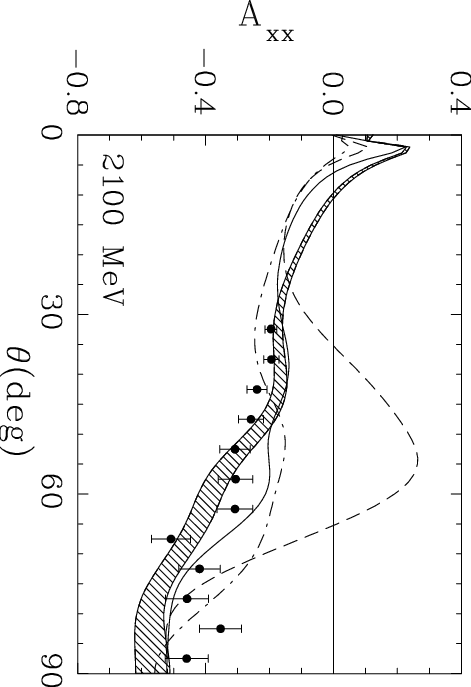}\hfill
\includegraphics[height=0.3\textwidth, angle=90]{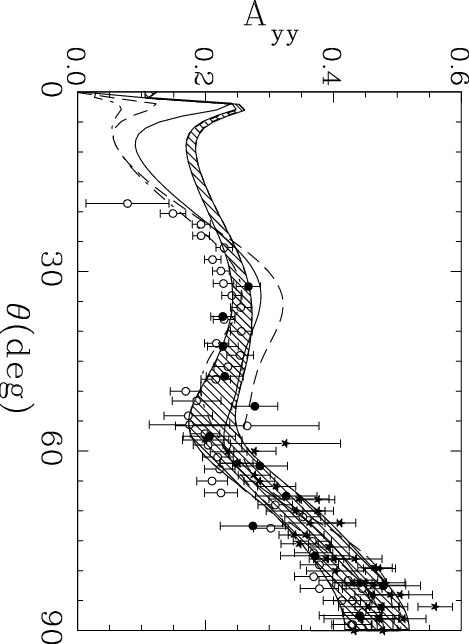}\hfill
\includegraphics[height=0.3\textwidth, angle=90]{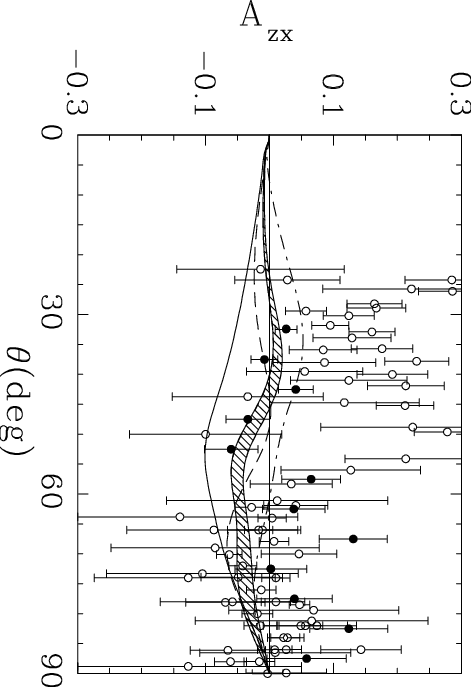}}
\vspace{10mm}
\caption{Angular distributions for $pp$
         elastic scattering at T$_p$ = 2100~MeV.
         The EDDA Collaboration data (filled
         circles) are from~\protect\cite{ba05}.
         Recent SATURNE~II $A_{yy}$
         measurements~\protect\cite{ag00,al01}
         are shown as stars.  Other previous
         measurements (for references see SAID
         database~\protect\cite{SAID}) within a
         10~MeV bin are shown as open circles.
         The SP07 (SP00~\protect\cite{sp00})
         solution is plotted as a solid (dashed)
         curve.  Saclay direct-reconstruction
         results~\protect\cite{lehar1} are shown
         as dash-dotted lines.  The GW
         single-energy solution is given by a 
         shaded band. \label{fig:g9}}
\end{figure*}

Some representative plots of the $A_{xx}$, $A_{yy}$ and
$A_{zx}$ data are given in Figs.~\ref{fig:g8} and 
\ref{fig:g9}.  The SP00 solution fails to 
correctly predict the EDDA $A_{xx}$ data above 1.5~GeV, 
and this discrepancy motivated the present study. The 
revised solution SP07 provides a much improved fit to 
these data. 

\begin{table}[th]
\caption{Comparison of present SP07 and previous
         SP00~\protect\cite{sp00}, 
         SM97~\protect\cite{sm97},
         SM94~\protect\cite{sm94},
         FA91~\protect\cite{fa91},
         SM86~\protect\cite{sm86}, and
         SP82~\protect\cite{sp82} energy-dependent
         partial-wave analyses.  The $\chi^2$ values
         for the previous solutions correspond to 
         our published results. \label{tab:tbl3}}
\vspace{2mm}
\begin{tabular}{|c|c|c|c|c|}
\colrule
Solution & Range & $\chi^2$/$pp$ data & Range & $\chi^2$/$np$ data \\
         & (MeV) &                  & (MeV) &                  \\
\colrule
SP07     &0--3000&    44463/24916   &0--1300&    21496/12693   \\
SP00     &0--3000&    36617/21796   &0--1300&    18693/11472   \\
SM97     &0--2500&    28686/16994   &0--1300&    17437/10854   \\
SM94     &0--1600&    22371/12838   &0--1300&    17516/10918   \\
FA91     &0--1600&    20600/11880   &0--1100&    13711/ 7572   \\
SM86     &0--1200&    11900/ 7223   &0--1100&     8871/ 5474   \\
SP82     &0--1200&     9199/ 5207   &0--1100&     9103/ 5283   \\
\colrule
\end{tabular}
\end{table}
\begin{table}[th]
\caption{Comparison of $\chi^2$/data for normalized
         (Norm) and unnormalized (Unnorm) $pp$ elastic 
         scattering data for the present SP07 and 
         previous SP00~\protect\cite{sp00} 
         solutions. \label{tab:tbl4}}
\vspace{2mm}
\begin{tabular}{|c|c|c|}
\colrule
 Range    &    SP07    &   SP00     \\
 (MeV)    & Norm/Unnorm& Norm/Unnorm\\
\colrule
   0-- 500& 1.4 /  4.5 & 1.3 /  4.3 \\
 500--1000& 1.4 /  9.9 & 1.3 /  8.9 \\
1000--1500& 2.0 /  6.7 & 2.2 /  6.3 \\
1500--2000& 2.0 /  7.0 & 2.6 /  6.4 \\
2000--2500& 3.0 /  8.3 & 3.7 /  8.1 \\
2500--3000& 3.3 / 29.6 & 3.6 / 50.5 \\
\colrule
\end{tabular}
\end{table}


\section{Summary and Conclusions}
\label{sec:conc}

We have generated a new fit to the full database of $pp$ and
$np$ elastic scattering data to 3~GeV (1.3~GeV for $np$ data).
This updated PWA provides a much improved fit to recent
polarized data measured by the EDDA collaboration at COSY. 
The new fit (SP07) has resolved some ambiguities found in 
comparing the previous SP00 energy-dependent and single-energy 
fits to DAR results from Saclay.  However, the resulting 
partial-wave amplitudes, in particular $^1D_2$ and $^1G_4$, 
have changed dramatically above 1~GeV. Given the impact of 
these data, and their absence above 2.5~GeV, our solution 
should be considered at best qualitative between 2.5 and 
3~GeV. 

Our agreement with the Saclay DAR amplitudes does not
imply agreement is necessary at the PWA level. As we have
noted, a PWA requires some model input, whereas the DAR
method requires only precise experimental data. As a test,
we generated a second solution having $^1D_2$ and $^1G_4$
amplitudes initially set to the SP00 values. After 
fitting data, the original SP07 behavior was regained. 
A fit having the SP00 behavior for $^1D_2$ and $^1G_4$
(a second $\chi^2$ minimum) was not found.

Further progress on $np$ scattering will require a program
to extend measurements above 1.3~GeV. Such a program has
been proposed for the Nuclotron at JINR, Dubna using a 
polarized deuteron beam and polarized proton 
target~\cite{dubna}.

\acknowledgments

The authors express their gratitude to C.~E.~Allgower,
F.~Bauer, J.~Blomgren. A.~D.~Carlson, J.~Franz,
F.~Hinterberger, H.~Lacker, A.~B.~Laptev, F.~Lehar, 
P.~Mermod, N.~Olsson, W.~Scobel, V.~I.~Sharov, H.~Spinka, 
W.~Tornow, S.~Vigdor, and V.~G.~Vovchenko providing 
experimental data prior to publication or for clarification 
of information already published.  This work was supported 
in part by the U.~S.~Department of Energy under Grant 
DE--FG02--99ER41110.  The authors acknowledge partial 
support from Jefferson Lab, by the Southeastern Universities 
Research Association under DOE contract DE--AC05--84ER40150.

\clearpage

\end{document}